%% file: paper.tex
\documentclass[runningheads]{llncs}
\usepackage{graphicx}
\begin{document}
\title{``That's our game!" : Reflections on co-designing a robotic game with neurodiverse children}
\titlerunning{``That's our game!" : Co-designing a game with neurodiverse children}
%
\author{Patricia Piedade \inst{1}\orcidID{0000-0002-7349-0237} \and
Isabel Neto\inst{2}\orcidID{0000-0003-2515-0446} \and
Ana Pires\inst{1}\orcidID{0000-0001-7747-7112} \and
Rui Prada\inst{2}\orcidID{0000-0002-5370-1893} \and
Hugo Nicolau\inst{1}\orcidID{0000-0002-8176-7638}}
\authorrunning{P. Piedade al.}
%
\institute{Interactive Technologies Institute, \\ Instituto Superior Técnico, University of Lisbon, Portugal \\ \and
INESC-ID, \\ Instituto Superior Técnico, University of Lisbon, Portugal \\ 
\email{\{patricia.piedade, isabel.neto, rui.prada, hugo.nicolau\}@tecnico.ulisboa.pt}}
\maketitle              
\begin{figure}[h] \
\includegraphics[width = \linewidth]{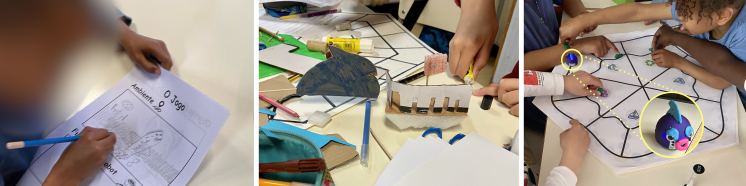}
\caption{Stages of the co-design process of a robotic game with neurodiverse children.}
\end{figure}
\label{fig:teaser}
\begin{abstract}
Many neurodivergent (ND) children are integrated into mainstream schools alongside their neurotypical  (NT) peers. However, they often face social exclusion, which may have lifelong effects. Inclusive play activities can be a strong driver of inclusion. Unfortunately, games designed for the specific needs of neurodiverse groups, those that include neurodivergent and neurotypical individuals, are scarce. 
Given the potential of robots as engaging devices, we led a 6-month co-design process to build an inclusive and entertaining robotic game for neurodiverse classrooms. We first interviewed neurodivergent adults and educators to identify the barriers and facilitators for including neurodivergent children in mainstream classrooms. Then, we conducted five co-design sessions, engaging four neurodiverse classrooms with 81 children (19 neurodivergent).
We present a reflection on our co-design process and the resulting robotic game through the lens of Self-Determination Theory, discussing how our methodology supported the intrinsic motivations of neurodivergent children.

\end{abstract}
%
%
%

\input{sections/1-Introduction}

\input{sections/2-CoDesign}

\input{sections/3-Reflection}
\input{sections/4-Conclusion}

\subsubsection*{Acknowledgements}
This work was supported by the European project DCitizens: Fostering Digital Civics Research and Innovation in Lisbon (GA 101079116), by the Portuguese Recovery and Resilience Program (PRR), IAPMEI/ANI/FCT under Agenda C645022399-00000057 (eGamesLab) and the Foundation for Science and Technology (FCT) funds SFRH/BD/06452/2021 and UIDB/50021/2020.

%
%
%
\bibliographystyle{splncs04}
\bibliography{biblio}
\end{document}

%% file: sections/1-Introduction.tex
\section{Introduction}
Play is an essential factor for childhood development \cite{hirsh-pasek,liu_hirsh_neuroscience}, aiding in the development of creativity, social skills and perception \cite{Fromberg1990,Fromberg1992,garvey1990,huizinga2014homo,Fromberg2012,peds}. Moments of play are a source of fun and a space for self-expression and exploration \cite{Fromberg2012,Johnson1987}. In fact, the United Nations Convention on the Rights of the Child recognises play as a fundamental right \cite{unicef}. Games, as a form of play, promote pleasurable engagement and players' well-being \cite{Iacovides2019,Jones2014}. Furthermore, games have the potential to promote inclusive and engaging experiences in mixed-ability scenarios \cite{inspo,map1,ball,ip6}.
However, neurodivergent children still face significant barriers regarding access to inclusive play scenarios and their above-mentioned benefits \cite{prop3,morris2023}. In this work, we take on the framework of the identity model of disability, using the concept of \textit{neurodiversity} to encompass the multitude of neurological differences in human brains.  Where most brains are \textit{neurotypical}, and some diverge from these norms, thus, referred to as \textit{neurodivergent} (e.g.: ADHD, autism, dyslexia, and intellectual disabilities) \cite{Dalton2013}. 

In a 2021 critical review of games developed by the HCI research community for neurodivergent players \cite{prop3}, Spiel and Gerling analysed 66 publications under the lens of Disability Studies and Self-Determination Theory. The authors conclude that serious games, designed for medical and training purposes, comprise most of the corpus. These games attempt to create an engaging facade for boring or repetitive tasks, tendentially prioritising training over enjoyment, and are driven by motivators outside neurodivergent interests. Furthermore, these games are often designed top-down, excluding the player from the design process and focusing on single-player dynamics, reducing opportunities for inclusive play and social interaction.

Though HRI is a growing field within HCI research, none of the games analysed by Spiel and Gerling \cite{prop3} included robots as game elements. Previous works regarding mixed-ability gaming have successfully leveraged robotic devices as proponents for inclusive play \cite{inspo,TACTOPI}. Moreover, outside the framework of games research, robots have proved to be a viable tool to create engaging experiences for neurodivergent individuals \cite{nrg6,nrg1,Knight2019,nci6}. Hence, there is unexplored potential for including robots in games geared towards neurodiverse groups.

Players' motivation is a central aspect of game design. Engaging gaming experiences require a motivated player \cite{SDTGames}. With wide use within HCI Games research \cite{SDTGames}, Self-Determination Theory (SDT) is a theory that models human motivation \cite{SDT2017,SDT2022} and the basis of Spiel and Gerling´s critical review of HCI games for neurodivergent players \cite{prop3}. SDT proposes three basic psychological needs that an activity must fulfil to promote intrinsic motivation: autonomy, competence and relatedness \cite{SDT2017,SDT2022}. Autonomy pertains to an individual's ability to choose their actions and circumstances according to their values and preferences \cite{SDT2017}. Competence describes a feeling of mastery over a particular subject and being met with appropriate challenges \cite{SDT2017}. Finally, relatedness is a feeling of social connectedness, being part of a group where one is cared for and cares for others through significant contributions \cite{SDT2017}. SDT argues that when an activity meets these three basic needs, it promotes motivation, which can lead to personal fulfilment and well-\cite{SDT2017}. Therefore, it is imperative that we take such needs into account when designing user experiences, such as games, or even participatory design processes.
Given the lack of games designed for a neurodiverse context and the potential of robots as game elements within this context, we set out to co-design a robotic game with and for neurodiverse classrooms. Throughout our co-design process, we aimed to centre neurodivergent interests and fill the research gap identified by Spiel and Gerling \cite{prop3}. In this paper, we reflect upon our co-design process and resulting game through the lens of SDT, critically evaluating our process as a form of accountability and informing future research within this context on how to better support self-determination within neurodiverse groups.

%% file: sections/2-CoDesign.tex
\section{Co-Design Process}

Aiming to bridge the gap within neurodiverse elementary school classrooms, we engaged in a multiple-methods co-design process by involving various stakeholders. 
Before engaging directly with the children, we engaged educators and neurodivergent adults in formative studies to better understand the barriers and facilitators to inclusion in a neurodiverse classroom.

\textbf{Co-Design Workshops.}
We proceeded to the co-design workshops within neurodiverse classrooms (Fig. \ref{fig:teaser}). We held these workshops at a local mainstream public elementary school. Four classrooms, two 2nd and two 4th grades, participated in the sessions. There were 81 students, aged 6 to 12, 19 of whom were neurodivergent (13 learning differences - G01ND3, G02ND1, G02ND6, G03ND3, G03ND4, G06ND1, G10ND5, G11ND3, G12ND1, G12ND3, G15ND2, G16ND1 and G16ND6, one dyslexia - G03ND4, two intellectual disabilities - G05ND1 and G05ND4, two ADHD - G06ND2 and G06ND3, one Down's Syndrome - G11ND5, and one Global Developmental Delay - G13ND1)\footnote{each child within this project is represented by an id G\textit{XX}\textit{NN}\textit{Y}, where \textit{XX} is a group number, \textit{NN} indicates if a child is ND or NT, and \textit{Y} is an in-group identifier.}. 

Over the course of four months, we conducted five hour-long sessions with each class. Teachers divided their classrooms into groups of 4 to 6 students based on usual seating arrangements, interests, and friendships. Throughout the process, children kept a project portfolio to store worksheets, drawings, and other design artifacts. We chose the Ozbot Evo \cite{ozobot_2022} as the robotic game element due to its target age range and proven efficacy in mixed-ability settings \cite{inspo} and with neurodivergent children \cite{Knight2019}. Each session started with a participative recap, where a researcher would prompt the children to recall events from past sessions. 
The first two sessions focused on building rapport with the children and familiarizing them with the robots. Children customized a folder to use as a project portfolio, decorated an Ozobot and partook in game-like activities to explore its features.
Session three focused on game design elements. Using Expanded Proxy Design \cite{prop1} and worksheets detailing essential game elements, children were asked to design games, themed around Oceans and Sustainability (curricular themes suggested by the teacher) for a stuffed animal friend with neurodivergent characteristics. Afterwards, we analysed the children's game concepts, identifying prominent game mechanics and themes and establishing the basic characteristics of our co-designed game. The final game concept consisted of a game of tag, where an Ozobot would chase players around a game board while the players attempted to complete mini-games to earn the most tokens and win the overall game.
For the last two sessions, each group of children formalized a concept for a mini-game, prototyped it, and play-tested it. Each group was given one of four themes inspired by their creations in session three and the two curricular themes proposed by the teachers. Researchers provided them with worksheets detailing game mechanic elements and crafting materials to actualize their ideas. Most mini-games generated had a rich narrative but vague rules.

\textbf{Game Design Process.}
Following the end of the co-design sessions, we conducted an iterative game design process supported by the results of the co-design workshops, culminating in a final prototype. The game, entitled ``The Shark Escape", was based around a classic ``tag" mechanic (as this was the most popular among children's prototypes) where players moved animal shaped pieces around a gameboard, evading being caught by the Ozobot, decorated like a shark. To avoid frustration related to waiting for one's turn, all players move at once, according to an automatic digital dice. Each player attempted to gather the three coloured tokens needed to return to their start position and win the game. 
To win tokens, players must land on mini-game spaces and win the corresponding mini-games: (1) Recycling - a two-player finger-football-style game in which players attempted to score goals with small coloured styrofoam balls in the correct recycling bin; (2) Treasure - a single-player game in which those not playing placed fish figurines on a grid, and the player attempted to move the Ozobot with the Ozobot Evo app \cite{EvoApp} remote control to reach the treasure without touching the fishes; (3) Animals - a classic multi-player memory game enhanced with AR, mapping  the cards to opensource 3D models through the Halo AR app \cite{HaloApp},  in which players attempt to find the most pairs of marine animals.
Winning a mini-game earned a player a corresponding token and a spin of the lucky prize wheel, which could earn them an extra reward (eg., the ability to move extra spaces). If caught by the shark (i.e. having their pawn knocked down by the Ozobot), players lose one token.

\textbf{Game Evaluation.}
To evaluate our prototype, we conducted a play-test session in neurodiverse classrooms. We recruited the four classrooms who had participated in the co-design sessions and an additional class as a control group. In total, 100 students, 26 of which were neurodivergent, tested the game. Classrooms were once again split into groups of 4 to 6 children, and each group played the game for one hour, while a researcher facilitated gameplay and observed.

%% file: sections/3-Reflection.tex
\section{Findings/Analysis through Self-Determination Theory}
Individual motivation is often disregarded when designing games for neurodivergent players \cite{prop3}. As a form of self-accountability, we analyse findings from our co-design process and game evaluation session under the lens of SDT. We focus this analysis on findings related to neurodivergent children, aiming to understand which practices best supported their self-determination.

\textbf{Competence.}
Taking into account the educational setting in which we situated our design process, competence was a key aspect to balance when creating co-design activities. 
Crafting activities activities presented manageable and fulfilling challenges. For instance, during session five, G06ND2 created a detailed boat structure and G02ND6 diligently coloured a gameboard prototype, both showed pride in their work and received praise from group-mates. On the other hand, less engaging group decision-making activities, such as conceptualizing games, proved frustrating for some. For example, G05ND4 often disengaged from the activities, G06ND2 frequently stood up so see what other groups were doing, G03ND4 struggled to complete the game elements worksheets, and G15ND2 struggled to have opinions heard. Strategies, such as encouraging children to draw their ideas (G05ND4), encouraging consensus rather than a majority vote (G06ND2), reminding children they could draw rather than write (G03ND4), and making turn-taking mechanics explicit (G15ND2), promoted neurodivergent children's sense of competence in these less entertaining activities. During our last visit to the school, one neurodiverse pair (G06ND1 and G05NT2) shared with us they planned on taking the knowledge they acquired to create their own game, indicating they felt competent in the game design knowledge aquired through the process.
Regarding the final prototype, we found that it provided enough of a challenge to keep the groups engaged while allowing everyone to succeed at a similar rate. Some neurodivergent children (eg., G05ND1, G05ND4, and G16ND6) struggled with counting spaces on the game board; however, they did not seem to perceive this as a lack of competence, simply moving in alternative ways around the gameboard and disregarding the dice.

\textbf{Relateness.}
Group work, especially within a school context, promotes socialisation, but not necessarily relatedness. We attempted to mitigate this issue by allowing teachers to form groups based on friendships and interests rather than enforcing a balance regarding the gender or neurodivergence of students within each group. We found that groups grew closer and learned to accomudate each other throughout the process. For instance, G02ND6's disruptive behaviour was initially perceived as bad, but, through the Expanded Proxy Design \cite{prop1} activity, they found a positive framework to employ it: their game concept consisted on pranking polluting humans out of wildlife habitats. This activity also allowed for self advocacy, with G05ND1 proudly stating: ``[The proxy] is like me! [...] She may not be able to read and write, but she has a good heart.''.
Regarding the game, we designed it to fit children's preference for competitive games (which most considered favorites due to being able to showcase competence), rather than inforcing socialization through a collaboration/cooperation mechanic.
However, the presence of a common enemy - the Ozobot -led children to spontaneously collaborate. For example, G06ND1 and G06ND3 encouraged G06ND2 to find matching pairs in the memory game, and G16ND1 shared his extra tokens with group-mates. 

\textbf{Autonomy.}
Once again, the school and group work settings are not natural promoters of autonomy as tasks are often dictated. Furthermore, the context of neurodiverse groups having to make joint decisions can lead to neurodivergent interests being overshadowed by the neurotypical majority. We aimed to reduce this issue by emphasizing that group decisions should be based on overall agreement rather than a majority vote. For example, G06ND2 felt very strongly about his ideas being include in the group's game, leading the rest of the group to find a way to incorporate everyone's contributions (a game where the player would have to sequentially complete various mini-games). We found that activities that required the creation of multiple design artifacts promoted individual autonomy. For instance, during session 4, G05ND1 created the mini-game's narrative, while G05ND2 drew the different scenes within it, adding specific details.
During the final play-test, children often bent the rules, which was accepted by their fellow group-mates. For example, whenever a group-mate landed on a mini-game spot G02ND6 would move his pawn there to play it as well, and lacking a proper place to store his token's G05ND2 started placing them on the Ozobot's plasticine shark fin. Players felt autonomous enough to play as they wished, we attribute this to a sense of ownership over the co-designed game. Still, the sit-down nature of the game left G06ND2 unfulfilled, seeking entertainment in unused game pieces, while other's played mini-games.

%% file: sections/4-Conclusion.tex
\section{Conclusion}
Having identified a need for gaming experiences designed for neurodiverse groups and the potential of robots to promote engagement, we set out to co-design a game with mainstream classrooms. We successfully co-designed a robotic board game with four neurodiverse classes, with a total of 80 students, 19 of which neurodivergent. We present a critical review of our design process under the lens of self-determination theory. Overall, we found that more entertaining activities with multiple resulting artifacts promototed promoted neurodivergent self-determination within the co-design process.
And the game's common enemy and allowance for rule-bending motivated neurodivergent children during gameplay. However, group decision-making activities, and the sit-down nature of the board-game require reworking in order to better advocate for neurodivergent competence, relatedness and autonomy.

In future work, we aim to further explore how the sense of ownership provided by the co-design process can promote autonomy in gameplay, and relatedness and competence in neurodiverse groups.